\documentclass{aastex}
\usepackage{emulateapj5}
\usepackage{apjfonts}

\begin{document}

\title{The Multiphase Intracluster Medium in Galaxy Groups Probed 
by the Ly$\alpha$ Forest}

\author{Hu Zhan and Li-Zhi Fang}
\affil{Department of Physics, University of Arizona, Tucson, AZ 85721}
\email{zhanhu@physics.arizona.edu, fanglz@physics.arizona.edu}

\author{\and David Burstein}
\affil{Department of Physics and Astronomy,
        Arizona State University, Tempe, AZ 85287-1504}
\email{burstein@samuri.la.asu.edu}

\begin{abstract}

The case is made that the intracluster medium (ICM) in present-day 
spiral-rich galaxy groups probably has undergone much slower evolution 
than that in elliptical-rich groups and clusters. The environments 
of protoclusters and protogroups at $z>2$ are likely similar 
to spiral-rich group environments at lower redshift. Therefore, 
like the ICM in spiral-rich groups today, the ICM in protogroups
and protoclusters at $z>2$ is predicted to be significantly 
multiphased. The QSO Ly$\alpha$ forest in the vicinity of galaxies 
is an effective probe of the ICM at a wide range of redshift.  
Two recent observations of Ly$\alpha$ absorption around 
galaxies by Adelberger et al. and by Pascarelle et al are reconciled,
and it is shown that observations support the multiphase ICM 
scenario. Galaxy redshifts must be very accurate for such studies
to succeed.  This scenario can also explain the lower metallicity 
and lower hot gas fraction in groups.

\end{abstract}


\section{Introduction}

Hot X-ray gas (with temperatures of $\sim 10^6$--$10^8$ keV) is found
in all galaxy clusters and groups in which it can currently be
detected \citep[see][hereafter M00]{bb02,m00}.  The fact that velocity
dispersions of groups and clusters correlate well with X-ray
temperature has shown that galaxy groups are distinct physical 
entities, not transient systems \citep[M00;][]{m03}.  In this
paper this hot gas is referred to as the intracluster medium (ICM),
with the understanding that this term also includes galaxy groups 
in its definition.

Cooling times for hot gas is a trivariate function of its
temperature, metallicity, and density \citep{sd93}.  As such, in
present-day galaxy groups and clusters, hot gas turns into cooler,
more neutral gas in at least two ways. In the centers of clusters, the
high density and high metallicity of the hot gas combine to predict
relatively short cooling times \citep[100's of millions of years;]
[]{f02}, while in the less dense outer parts of clusters, the
cooling times are much longer.  Separately, in groups with low
velocity dispersion \citep[i.e., those dominated by spiral and
irregular galaxies;][]{m96, mz98, zm98}, the still relatively
metal-rich hot gas can become neutral quickly because of its lower 
temperature and interaction with neutral hydrogen in the galaxies 
in these groups \citep{bb02}. Thus the ICM of spiral-rich galaxy 
groups today (e.g. the Local Group) can be considered to be multiphase.

The question then is how multiphase is the ICM at high redshift?
Answering this question is crucial to understanding the evolution of gas
in galaxy groups and clusters. In this paper the case is made
that a multiphase ICM exists in high-$z$ galaxy groups and clusters 
as they are forming (\S~\ref{sec:m-phase}). The evidence for this 
conjecture that comes from the Ly$\alpha$ forest is provided in 
\S~\ref{sec:lyman}, where it is shown how to reconcile two 
recent observations that somewhat contradict each other.  
Further implications of the multiphase ICM are discussed in 
\S~\ref{sec:con}.  An $\Omega = 0.3$, $\Lambda = 0.7$ universe is
assumed in this paper, and the distance scales quoted below are all
comoving.

\section{A Likely Multiphase ICM at High Redshift}
\label{sec:m-phase}

From present-day hot ICM gas, two things must be true when groups 
and clusters are forming: First, because all hot ICM gas has a 
significant abundance of metals (0.1--0.5 times solar, M00), much
of this hot gas had to have seen the interiors of stars.
Second, because of the strong correlation of group/cluster velocity
dispersion with X-ray temperature (M00), the gas became hot through
heating by supernovae and stirring by the virial motion of the
galaxies.  This observational picture is consistent with the current
hierarchical, clustering, merging scenario, which predicts that
today's galaxies formed from smaller objects in an environment
dominated by cold dark matter \citep[e.g.][]{nfw95}.

From studies of galaxy groups and clusters in the nearby universe
\citep[see the Nearby Galaxy Catalog by][]{t88}, the vast majority of
nearby galaxies are well-associated with either a galaxy group or a
galaxy cluster.  Of the nearly 2450 galaxies within 3000 km s$^{-1}$
distance of our Local Group that are in the Nearby Galaxy Catalog, only
35 are not put in a group, cloud, or cluster. And, most of these 35
galaxies are within less than 1 Mpc of a group or cluster.
Furthermore, the majority of galaxies in the Nearby Galaxy Catalog are
in galaxy groups (most of which are low-velocity dispersion,
spiral-rich groups), not in elliptical-dominated clusters such as the
Virgo cluster.  As such, it is reasonable to assume that if galaxies
are seen at high redshift, they are more likely to be part of a group
(even though groups are hard to identify at high-$z$).

When galaxies are in the beginning throes of formation, the ICM should have
similar properties to the diffuse $\sim 10^4$ K intergalactic medium
(IGM), whose temperature scales with the density as $T \propto
\rho^\alpha$, with $\alpha=0.3$--$0.6$ \citep{hg97}. At the start of
galaxy formation (as well as group and cluster formation) the
density of the ICM may not be much greater than the cosmic mean;
hence the ICM cannot be initially much hotter than $10^4$ K.
At this stage, the ICM is relatively uniform in the gravitational well.

The ICM will become hotter as galaxies evolve and inject hot gas and
energy into it. However, the heating before $z = 2$ is not strong
enough to heat the ICM to $10^6$--$10^8$ K globally.  Supernova
heating is proportional to the star formation rate, which peaks at $z
\simeq 1-2$ \citep{s99}. Virial shocks are not likely to happen at $z
> 2$ or in systems less massive than $10^{11}M_{\sun}$
\citep{bd03}. Other sources of heating are active galactic nuclei
(AGNs), whose number density increases sharply toward $z \simeq 2$,
but does not evolve much beyond $z \simeq 2$ \citep{mhs00}. Yet,
recent observations around protoclusters \citep[][hereafter A03]{a03}
imply that AGN heating has not drastically affected the ICM at
$z\simeq 3$.  Therefore, AGNs may not have a primary role in heating
the ICM in all clusters and groups.

On the other hand, cooling of gas at its higher density at higher
redshift is relatively quick. This is supported by the argument that
metallicity of this gas had to have been seeded quickly by the
first generations of star formation. As a result, the metallicity of the 
high-$z$ ICM hot gas probably is not too much lower than it is today
It is therefore reasonable to assume that pockets of moderately-metal-rich, 
$10^4$--$10^5$ K gas exist in $z > 2$ protogroups and protoclusters.

The relatively weak heating and quick cooling necessarily lead to a
widely distributed, multiphase ICM both in protoclusters and in
protogroups at high redshift.  Spiral-rich groups can maintain such a 
distributed, multiphase ICM since they receive less energy input 
from fewer galaxies, and since many of their galaxies have neutral gas 
with which the cooler hot gas can interact \citep{bb02}. In other 
words, the ICM in spiral-rich groups today probably has undergone 
a much slower evolution than the ICM in rich clusters and
elliptical-rich galaxy groups.  With this scenario, 
the environments of protoclusters and protogroups at 
$z>2$ are expected to be similar to those in present-day spiral-rich 
galaxy groups, and therefore the ICM in all galaxy 
systems is multiphase at $z>2$.

In contrast, present-day rich clusters and elliptical-rich groups cannot 
preserve such a distributed multiphase ICM since their member galaxies 
supply large amounts of virial energy to the ICM. This picture is 
consistent with the fact that rich clusters are mostly observed at 
$z<1$ \citep{h92}, and the X-ray emitting ICM in clusters and 
groups has to be mostly formed between $z = 2$ and today.

\section{The Multiphase ICM Probed by the Ly$\alpha$ Forest}
\label{sec:lyman}

The above scenario predicts that foreground galaxies, through a
strongly multiphased ICM that contains a large number of strong
Ly$\alpha$ absorbers, can influence the Ly$\alpha$ forest of a
background QSO. Accordingly one expects to see a higher number 
of strong Ly$\alpha$ absorption lines and a lower mean 
Ly$\alpha$ flux at smaller impact parameters between foreground 
galaxies and background QSOs.  This gives us a practical criterion: 
an ICM is strongly multiphased if it contains stronger fluctuations 
(measured by Ly$\alpha$ absorptions) than the IGM.

\subsection{Observations} \label{sec:obs}

This scenario is consistent with a number of observations. \citet{l95},
\citet{c98}, and \citet[][hereafter P01]{p01} find that galaxies that
are not influenced by the QSOs (i.e, the ``far sample'' of galaxies in
P01) which are at impact parameters $\rho \leq 180\ h^{-1}$kpc are
much more often associated with Ly$\alpha$ absorption in the
background QSO spectrum than those at $\rho > 180\ h^{-1}$kpc.  The
rest frame equivalent widths $W$ of absorption tend to increase as
$\rho$ decreases from $1\ h^{-1}$Mpc to $10\ h^{-1}$kpc (P01, Fig.~1).

It is conservative to assume that most galaxies in the far
sample are in groups (the same as they are in the local universe).
The line statistics thus suggest that the ICM in groups does contain
more strong Ly$\alpha$ absorbers than the diffuse IGM.  The far sample
in P01 ranges from $z=0.02$ to $1$, over which groups are seen to
contain sub-keV gas \citep[M00;][]{xw00}, consistent with a
multiphase ICM.  On the other hand, 
Ly$\alpha$ lines around galaxies that are subject to QSO influence 
are almost always weaker than those in the far sample, and they do not 
depend much on the impact parameter (P01). This shows that the ICM 
cannot remain multiphase under the influence of QSOs, and 
consequently AGNs may be responsible for heating the ICM in some 
groups and clusters at $z \lesssim 1$.

At a higher redshift, $z\simeq 3$, A03 show that mean Ly$\alpha$
transmissivity generally increases with increasing impact parameter
between foreground Lyman-break galaxies (LBGs) and background QSOs.
It reaches cosmic mean transmissivity of 0.67 at $\rho \gtrsim 6\
h^{-1}$ Mpc. This result indicates that the ICM at $z\simeq3$
has a higher overall neutral fraction than the IGM. Since \ion{H}{1} 
concentrates where the Ly$\alpha$ lines are, the Ly$\alpha$ lines 
near the LBGs must have substentially larger equivalent widths than 
the Ly$\alpha$ forest lines. This is consistent with both the line 
statistics above and a multiphase ICM.

LBGs are more likely to be massive systems \citep{b98} that reside in 
dense regions, where groups and clusters are found today. This 
strengthens the argument that Ly$\alpha$ absorption associated with 
the LBGs out to 6 \mbox{$h^{-1}$ Mpc} is due mostly to the ICM.  
In fact, the LBGs in A03 are claimed to likely be in protoclusters, 
of which at least a few will evolve into rich clusters at $z = 0$ 
that contain most of their baryonic mass in X-ray emitting hot gas.

\subsection{The Discrepancy}

\begin{figure}
\epsscale{0.45}
\plotone{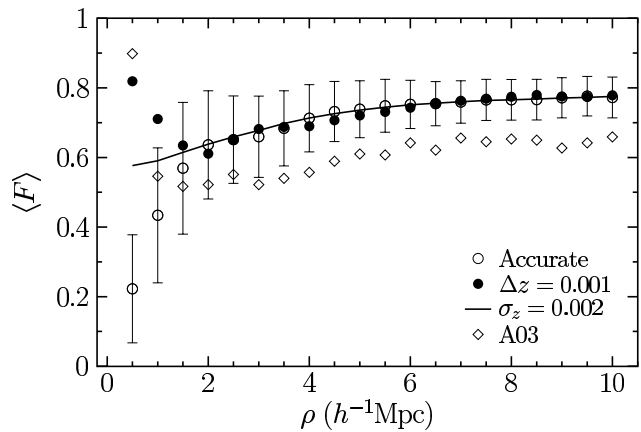}
\caption[f1.eps] {Mean Ly$\alpha$ flux around galaxies. The open circles,
filled circles, and the
solid line are the cases with no error in redshift, a systematic error
of $\Delta z = 0.001$, and a Gaussian error with dispersion
$\sigma_z=0.002$, respectively. The error bars of the three are similar,
but only plotted for the open circles for clarity. The diamonds are from
A03. The segment of the Ly$\alpha$ forest in our analysis is from
$z=2.16$ to $2.70$, so the cosmic mean flux is higher than that at $z=3$.
This leads to the difference between A03 and the rest of the data at large
impact parameters.
\label{fig:meanF}}
\end{figure}

Despite the general agreement, flux statistics in A03 are not consistent 
with line statistics at $\rho < 0.5\ h^{-1}$Mpc. An increase of mean 
flux from $0.55$ at $1\ h^{-1}$Mpc to $0.9$ at $0.5\ h^{-1}$Mpc 
(see Figure \ref{fig:meanF}) is reported in A03.

LBGs are strong UV sources, and the escape fraction of ionizing photons
from LBGs at $z=3$ is higer than that from local star burst galaxies
\citep{dbb01, spa01, gcd02}. It is possible that the rise of the mean flux
close to LBGs at $z=3$ is due to the photoionization by the LBGs themselves
\citep[similar to the proximity effect of QSOs, e.g.][]{bdo88, gcd96,
sbd00}, which is absent in low-redshift galaxies. One can roughly
estimate the size of influence by LBGs as follows. If the luminosity
of a typical QSO is $10^{46}$ \mbox{erg s$^{-1}$} and if it has a
line-of-sight (LOS) proximity zone of 36 $h^{-1}$Mpc (4000 \mbox{km
s$^{-1}$} at $z=3$), then an LBG with a luminosity of $10^{41}$ \mbox{erg
s$^{-1}$} would create a highly ionized (higher than the IGM) bubble
with a radius of at most 0.8 $h^{-1}$Mpc, where a proportionality
between volume and luminosity has been assumed. The radiation of QSOs
is most likely anisotropic, as suggested by null detections of
foreground QSO proximity effect \citep{c89, mk92, lw01}, so that the
relatively isotropic radiation from LBGs should considerablly reduce
its radius of influence. In addition, QSOs have harder spectra than
LBGs, which help to create larger proximity zones, owing to larger
fractions of high-engergy photons that can travel farther to ionize
\ion{H}{1} because of their smaller cross sections for photoionization. 
Therefore, an estimate of 0.8 $h^{-1}$Mpc is a conservative upper 
bound for the radius that can be affected by radiation from LBGs.  
Alternatively, A03 suggests that supernova-driven winds with a 
sustaining speed of 600 \mbox{km s$^{-1}$} for a few hundred million 
years may be able to deplete \ion{H}{1} and reduce Ly$\alpha$ 
absorption in a region of comparable size.

Regardless how the highly ionized bubble is created, however,
\citet{w03} show with hydrodynamical simulations that even if one 
removes all \ion{H}{1} within 1.5 $h^{-1}$Mpc from LBGs in real space, 
the mean flux at $\rho < 0.5\ h^{-1}$Mpc is still lower than 0.67. 
It is because the infalling \ion{H}{1} from large radii could be 
easily seen within $0.5\ h^{-1}$Mpc in redshift space. This is similar 
to finger of god effect, which smears real-space information along the 
LOS. For example, the Hubble constant at $z=3$ is 446$h$ 
\mbox{km s$^{-1}$Mpc$^{-1}$}, so that an infall speed of 120 
\mbox{km s$^{-1}$} will enable Ly$\alpha$ absorbers at a LOS distance
of 1.5 $h^{-1}$Mpc from an LBG in real space to appear less than
0.5 $h^{-1}$Mpc from the LBG in redshift space. Since the peculiar velocity
of a gas clump in a cluster easily exceeds 120 \mbox{km s$^{-1}$}, it will
completely erase the LBG bubble in redshift space.

In summary, the flux statistics found by by A03 are incompatable with
the line statistics found by P01 at $\rho < 0.5\ h^{-1}$Mpc. This difference
cannot be easily attributed to ionization by LBG themselves, through either
radiation or winds, unless LBGs could ionize an unrealistically large
volume (radius $\gg 1\ h^{-1}$Mpc) that cannot be erased by the virial
motions of gas clouds.

\subsection{Resolving the Discrepancy: Redshift Errors}

The mean flux at an impact parameter from a galaxy is meaningful 
only if it is calculated within a compatible distance from the 
galaxy along the LOS. The rms error of galaxy redshift is $\sigma_z=0.002$ 
in A03, which corresponds to 1.3 $h^{-1}$Mpc at $z \simeq
3$. In turn, redshift errors can lead to an inaccurate estimate of 
the mean flux within $\rho = 1.3\ h^{-1}$ Mpc. A03 evaluates the mean
flux in their data using the flux--galaxy correlation function 
to suppress the effect of redshift error. This method only helps 
if the sample is large. However, A03 has only 3 galaxies at 
$\rho< 0.5\ h^{-1}$Mpc without Ly$\beta$ contamination (3 more 
with Ly$\beta$ contamination), so redshift errors could still alter 
mean flux within $\rho = 1.3\ h^{-1}$Mpc.

The effect of a small redshift error $\Delta z$ on the mean flux
is examined using the Keck high resolution spectrum of QSO
HS1700+64 \citep[for observational details and data reduction techniques,
see][]{kt97}. This QSO is at $z = 2.74$, and the spectral resolution 
of the data used is 8 \mbox{km s$^{-1}$}. Only the segment between 
$z = 2.16$ and $2.70$ is analyzed in order to exclude Ly$\beta$ 
absorptions and the QSO proximity zone. In an idealized case that a 
galaxy is at the center of a Ly$\alpha$ line, mean flux of the 
Ly$\alpha$ forest around the galaxy redshift should increase with 
the scale over which this mean flux is calculated, until it reaches 
the cosmic mean.

A mock sample of 59 galaxies is made by assigning the positions of
strong Ly$\alpha$ lines ($4$ \AA{} $\geq W \geq 0.5$ \AA{}) to
galaxies. If two lines are closer than 2.5 \AA{}, the stronger one 
remains. The line width criterion is arbitrary, but it only 
contributes to the dispersions and details, not the general trend. 
Line widths are adopted from \citet{db96} with corrections 
to line positions. Impact parameters are not assigned; rather  
mean fluxes are calculated within the LOS distances (still reported as 
$\rho$) that are equal to the impact parameters as argued above.

Mean flux around accurate galaxy positions is shown in
Figure~\ref{fig:meanF} with open circles, which decreases
monotonically as $\rho$ decreases. A small systematic error of $\Delta
z = 0.001$ (filled circles) can easily raise the mean flux at $\rho
\leq 1.5\ h^{-1}$Mpc. If the error is Gaussian with a dispersion
$\sigma_z = 0.002$ (solid line), the mean flux drops slightly from
$\rho =2\ h^{-1}$Mpc to $\rho = 0.5\ h^{-1}$Mpc, but it never rises at
smaller $\rho$.

The same Gaussian error was tested by A03, but no significant effect 
was found. If the positions of the few galaxies in A03 at 
$\rho < 0.5\ h^{-1}$Mpc are all off by $\Delta z = 0.002$, one 
cannot recover a meaningful ensemble average of the mean flux by 
adding the Gaussian error $\sigma_z=0.002$ to their positions. 
The positional error of a small number of galaxies is simply not 
well-characterized by a Gaussian distribution. More observations 
are needed to conclusively establish the mean flux at small 
impact parameters.

{\it If} the deficit of Ly$\alpha$ absorption at $\rho < 0.5\
h^{-1}$Mpc is the result of errors in galaxy positions, then the
flux statistics and the line counting statistics give a
consistent picture of a multiphase ICM that exists in high-$z$ galaxy
agregations and low-$z$ groups. If not, one will have to explain how
LBGs ionize a sphere of radius $\gg 1\ h^{-1}$Mpc in real space.

\subsection{Neutral Hydrogen in the ICM at $z\simeq 3$}

Given the mean flux, $\langle F \rangle$, one can estimate the
density of neutral hydrogen $n_H$ using \citep{p93}
\begin{equation} \label{eq:nH}
n_H \simeq 2.4 \times 10^{-11} \Omega^{1/2} h (1+z)^{3/2}
\ln\langle F \rangle^{-1}.
\end{equation}
The \ion{H}{1} column density $N_H$ at $z=3$ is roughly
\begin{equation} \label{eq:NH}
N_H \simeq n_H l \simeq 8.1\times 10^{13} \ln\langle F \rangle^{-1}
\left(l/h^{-1}\mbox{Mpc}\right)\ \mbox{cm}^{-2},
\end{equation}
where $l$ is the (comoving) LOS path length through the ICM.
With the lowest mean flux, $\langle F \rangle=0.52$, near LBGs
in A03, and $l=12\ h^{-1}$Mpc for a spherical ICM that extends to
$6\ h^{-1}$Mpc (\S \ref{sec:obs}), equation (\ref{eq:nH}) 
yields an upper limit of
$N_H=6.4 \times 10^{14}$ cm$^{-2}$ in the ICM at $z=3$.
This is several orders of magnitude lower than Galactic values, 
so it can easily escape detection from 21 cm radio and other 
observations, even if it still exists in the ICM today.

\section{Discussion and Conclusions}
\label{sec:con}

The Ly$\alpha$ forest in the vicinity of foreground galaxies is 
potentially a great probe for studying the ICM. Caution must be taken, 
however, since knowing the accuracy of galaxy redshift is crucial when 
probing at small impact
parameters. Nevertheless, the available observations point to a 
multiphase ICM at $z = 3$, which exists partially in the form of strong 
Ly$\alpha$ absorbers. It is also evident from the Ly$\alpha$ forest 
that a multiphase ICM is preserved in spiral-rich galaxy groups 
today, consistent with what can be predicted about the cooling times 
for the hot gas in such galaxy groups \citep{bb02}. An independent 
piece of evidence comes from recent discovery of the warm-hot 
$10^5$--$10^7$ K intra{\it group} medium through observations of 
the high velocity \ion{O}{6} absorption clouds around our Galaxy at 
a likely distance $\gtrsim 100$ kpc \citep[e.g.,][]{n03}. For rich 
clusters, the ICM may have been homogenized by galaxies, so that 
it is unlikely to be multiphase except in core regions where 
cooling times are short, and therefore expected to show much less 
Ly$\alpha$ absorption than the IGM.

It is found that ionization by LBG themselves via radiation or
supernovae-driven winds cannot easily account for the lack of
Ly$\alpha$ absorption at $\rho < 0.5\ h^{-1}$Mpc in A03, because
LBGs do not have sufficient energy to ionize a region of size 
$\gg 1\ h^{-1}$Mpc in real space. Errors in galaxy positions can 
cause the apparent rise of mean flux at $\rho < 0.5\ h^{-1}$Mpc
found by A03.

The origin of the multiphase ICM is the competition between heating 
and cooling in these environments. The detailed properties of the
absorbers are not well determined by Ly$\alpha$ observations alone. 
This is because even though cool ($\sim 10^4$ K) and dense absorbers 
are common at $z=3$, shocked gas clumps appear more often at low 
redshift. These gas clumps can have high \ion{H}{1} column densities
\citep[e.g.][]{dhk99}, in spite of higher temperatures ($\gtrsim 10^5$ K).

The existence of a multiphase ICM requires effective cooling and low
heating. This is consistent with groups having lower fraction of mass
in hot gas than clusters, if a higher fraction of the ICM in groups is
kept in cool and warm phases. In addition, since galaxies heat the ICM
by feedback and motion, less heating means lower metallicity in the
ICM in galaxy groups, which is observed (M00).

A multiphase ICM may also contribute to the excess of entropy in the
cores of poor clusters and groups \citep{pcn99}, which requires lower
electron density than predicted by the hierarchical clustering model.
Efficient cooling in poor clusters and groups may reduce hot gas
density in a short time.  If this picture is correct, one might
circumvent the problem of missing soft X-ray emission encountered by
the cooling flow model \citep{f01, wfn01}, which lowers the hot gas
density by on-going cooling.

\acknowledgments
We thank D. Tytler for kindly providing the Keck spectrum HS1700+64,
and the referee for helpful comments.

\begin{thebibliography}{}

\bibitem[Adelberger et al.(2003)]{a03} Adelberger, K.L., Steidel, C.C.,
Shapley, A.E., Pettini, M. 2003, \apj, 584, 45, A03

\bibitem[Bajtlik, Duncan \& Ostriker(1988)]{bdo88} Bajtlik, S.,
Duncan, R.C., Ostriker, J.P. 1988, \apj, 327, 570

\bibitem[Baugh et al.(1998)]{b98} Baugh, C.M., Cole, S., Frenk, C.S.,
\& Lacey, C.G. 1998, \apj, 498, 504

\bibitem[Birnboim \& Dekel(2003)]{bd03} Birnboim, Y, \& Dekel, A. 2003,
submitted to \mnras, astro-ph/0302161

\bibitem[Burstein \& Blumenthal(2002)]{bb02} Burstein, D., \&
Blumenthal, G. 2002, \apj, 574, L17

\bibitem[Chen et al.(1998)]{c98} Chen, H.-W., Lanzetta, K.M.,
Webb, J.K., \& Barcons, X. 1998, \apj, 498, 77

\bibitem[Crotts(1989)]{c89} Crotts, A.P.S. 1989, \apj, 336, 550

\bibitem[Dav\'e et al.(1999)]{dhk99}  Dav\'e, R., Hernquist, L.,
Katz, N., \& Weinberg, D.H. 1999, \apj, 511, 521

\bibitem[Deharveng et al.(2001)]{dbb01} Deharveng, J.-M., Buat, V.,
Le Brun, V., Milliard, B., Kunth, D., Shull, J.M., \& Gry, C. 2001,
\aap, 375, 805

\bibitem[Dobrzycki \& Bechtold(1996)]{db96} Dobrzycki, A., \&
Bechtold, J. 1996, \apj, 457, 102

\bibitem[Fabian et al.(2001)]{f01} Fabian, A.C., Mushotzky, R.F.,
Nulsen, P.E.J., \& Peterson, J.R. 2001, \mnras, 321, L20

\bibitem[Fabian et al.(2002)]{f02} Fabian, A.C., Allen, S.W.,
Crawford, C.S., Johnstone, R.M., Morris, R.G., \& Sanders, J.S. 2002,
\mnras, 332, L50

\bibitem[Giallongo et al.(1996)]{gcd96} Giallongo, E., Cristiani, S.,
D'Odorico, S., Fontana, A., \& Savaglio, S. 1996, \apj, 446, 46

\bibitem[Giallongo et al.(2002)]{gcd02} Giallongo, E., Cristiani, S.,
D'Odorico, S., \& Fontana, A. 2002, \apj, 568, L9

\bibitem[Henry et al.(1992)]{h92} Henry, J.P., Gioia, I.M., Maccacaro, T.,
Morris, S.L., Stocke, J.T., \& Wolter, A. 1992, \apj, 386, 408

\bibitem[Hui \& Gnedin(1997)]{hg97} Hui, L., \&
Gnedin, N.Y. 1997, \mnras, 292, 27

\bibitem[Kirkman \& Tytler(1997)]{kt97} Kirkman, D., \& Tytler, D.
1997, \apj, 484, 672

\bibitem[Lanzetta et al.(1995)]{l95} Lanzetta, K.M., Bowen, D.V.,
Tytler, D., \& Webb, J.K. 1995, \apj, 442, 538

\bibitem[Liske \& Williger(2001)]{lw01} Liske, J., \& Williger, G.M.
2001, \mnras, 328, 653

\bibitem[Miyaji, Hasinger \& Schmidt(2000)]{mhs00} Miyaji, T., Hasinger,
G., \& Schmidt, M. 2000, \aap, 353, 25

\bibitem[M\o{}ller \& K\ae{}rgaard(1992)]{mk92} M\o{}ller, P.,
\& K\ae{}rgaard, P. 1992, \aap, 258, 234

\bibitem[Mulchaey(2000)]{m00} Mulchaey, J. S. 2000, \araa, 38, 289, M00

\bibitem[Mulchaey et al.(2003)]{m03} Mulchaey, J.S., Davis, D.S.,
Mushotzky, R.F., \& Burstein, D., 2003, \apjs, 145, 39

\bibitem[Mulchaey et al.(1996)]{m96} Mulchaey, J. S., Davis,
D. S., Mushotzky, R. F., \& Burstein, D. 1996, \apj, 456, 80

\bibitem[Mulchaey \& Zabludoff(1998)]{mz98} Mulchaey, J.S.,
\& Zabludoff, A. I. 1998, \apj, 496, 73

\bibitem[Navarro, Frenk \& White(1995)]{nfw95} Navarro, J.F., Frenk, C.S.,
\& White, S.D.M. 1995, \mnras, 275, 56

\bibitem[Nicastro et al.(2003)]{n03} Nicastro, F., Zezas, A., Elvis, M.,
Mathur, S., Fiore, F., Cecchi-Pestellini, C., Burke, D., Drake, J., \&
Casella, P. 2003, \nat, 421, 719

\bibitem[Pascarelle et al.(2001)]{p01} Pascarelle, S.M., Lanzetta, K.M.,
Chen, H.-W., \& Webb, J.K. 2001, \apj, 560, 101, P01

\bibitem[Peebles(1993)]{p93} Peebles, P.J.E. 1993,
{\it Principles of Physical Cosmology}
(Princeton: Princeton University Press)

\bibitem[Ponman, Cannon \& Navarro(1999)]{pcn99} Ponman, T.J.,
Cannon, D.B., \& Navarro, J.F. 1999, \nat, 397, 135

\bibitem[Scott et al.(2000)]{sbd00} Scott, J.,
Bechtold, J., Dobrzycki, A., \& Kulkarni, V.P. 2000, \apjs, 130, 67

\bibitem[Steidel, Pettini \& Adelberger(2001)]{spa01} Steidel, C.C.,
Pettini, M., \& Adelberger, K.L. 2001, \apj, 546, 665

\bibitem[Steidel et al.(1999)]{s99} Steidel, C.C., Adelberger, K.L.,
Giavalisco, M., Dickinson, M., \& Pettini, M. 1999, \apj, 519, 1

\bibitem[Sutherland \& Dopita(1993)]{sd93} Sutherland, R.S., \&
    Dopita, M.A., 1993, \apjs, 88, 253

\bibitem[Tully(1988)]{t88} Tully, R. B. 1988, Nearby Galaxies Catalog,
(New York: Cambridge Univ. Press)

\bibitem[Weinberg et al.(2003)]{w03} Weinberg, D.H., Dav\'e, R.,
Katz, N., \& Kollmeier, J.A. 2003, The Emergence of Cosmic Structure, 
Proceedings of the 13th Annual Astrophysics Conference in Maryland, 
eds. S. Holt and C. Reynolds (New York: AIP), in press (astro-ph/0301186)

\bibitem[Wu, Fabian \& Nulsen(2001)]{wfn01} Wu, K.K.S., Fabian, A.C.,
\& Nulsen, P.E.J. 2001, \mnras, 324, 95

\bibitem[Xue \& Wu(2000)]{xw00} Xue, Y.-J., \& Wu, X.-P. 2000,
\apj, 538, 65

\bibitem[Zabludoff \& Mulchaey(1998)]{zm98} Zabludoff, A.I.,
\& Mulchaey, J. S. 1998, \apj, 496, 39


\end{thebibliography}
\end{document}